\documentclass[aps,prl,showpacs,noshowkeys,amsmath,amssymb,amsfonts,reprint]{revtex4-1}
\usepackage{graphicx}
\usepackage{dcolumn}
\usepackage{bm}
\usepackage{color}
\usepackage{ulem} 

\begin{document}
\title{Clogging and Jamming of Colloidal Monolayers Driven Across a Disordered Landscape}
\author{Ralph L. Stoop$^1$}
\author{Pietro Tierno$^{1,2,3}$}
\email{ptierno@ub.edu}
\affiliation{
$^1$Departament de F\'isica de la Mat\`eria Condensada, Universitat de Barcelona, Barcelona, Spain.\\
$^2$Universitat de Barcelona Institute of Complex Systems (UBICS), Universitat de Barcelona, Barcelona, Spain\\
$^3$Institut de Nanoci\`encia i Nanotecnologia, IN$^2$UB, Universitat de Barcelona, Barcelona, Spain.}
\date{\today}
\begin{abstract}
We experimentally investigate the  
clogging and jamming of interacting paramagnetic colloids driven through 
a quenched disordered landscape of fixed obstacles.
When the particles are forced to cross 
a single aperture between two obstacles, 
we find an intermittent dynamics characterized by an exponential distribution of burst size. 
At the collective level, 
we observe that quenched disorder decreases 
the particle flow, but  
it also greatly enhances the 
"faster is slower" effect, that occurs  
when increasing the particle speed.
Further, we show that clogging events 
may be controlled 
by tuning the pair interactions between the 
particles during transport,
such that the colloidal flow decreases 
for repulsive interactions,
but increases for  
anisotropic attraction. 
We provide an experimental test-bed 
to investigate the crucial role 
of disorder on  
clogging and jamming in driven microscale matter.
\end{abstract}
\maketitle
Particle transport through heterogeneous media 
is a fundamental problem across several disciplines 
as physics, biology and engineering.
In condensed matter, 
the inevitable presence of quenched disorder 
affects the transport properties of several systems, from 
vortices in high $T_c$ superconductors~\cite{Bha93,Bla94},
to electrons on the surface of liquid helium~\cite{Ree12},
skyrmions~\cite{Rei15},
and active matter~\cite{Ale17}. 
At the macroscopic scale, 
disorder in form of obstacles, wells or barriers 
severely alters the flow of bubbles~\cite{Cha10,Dre17},
granular media~\cite{Bid93,To01}, bacteria~\cite{Zen16}, sheep~\cite{Gar15} or pedestrians~\cite{Yan09}. 
Already a collection of particles that are forced 
to pass through a small constriction displays a complex dynamics, including flow intermittency, 
a precursor of blockage via 
formation of particle bridges and arches. 
The latter general phenomenon is known as clogging, 
and is responsible for the flow arrest in different technological systems, 
from microfluidics, to silo discharge,
and gas and oil flow through pipelines.
Clogging is also directly related to jamming,
which occurs   
when, above a threshold density, 
a loose collection of elements 
reaches a solid-like phase with a finite yield~\cite{Cat98}. 
Jammed systems are associated with the existence of a
a well defined rigid state and 
a new type of zero-temperature critical point~\cite{Liu98,Goo16}.
In contrast,
the local and spatially heterogeneous nature 
of clogging makes this phenomenon more difficult to characterize and 
to control,
despite its technological relevance.

In previous experimental realizations, 
clogging has 
received much attention   
at the level of a single bottleneck~\cite{Ike14},
while studies addressing the dynamics of microscale systems driven through heterogeneous landscapes are rather scarce~\cite{Ale17}. 
In contrast, recent theoretical works demonstrated
the rich phenomenology 
of transport and clogging across ordered~\cite{Re14,Yan17} or 
disordered~\cite{Gla16,Ngu17} landscapes. 
The advantage of using colloidal particles as model systems for clogging  
is their flexibility, since external fields may be used to create driving forces for 
transport, or to tune {\it in situ} 
the pair interactions.

Here we experimentally
investigate the flow properties  of a monolayer of   
paramagnetic colloidal particles that 
is driven across a heterogeneous landscape composed 
of disordered obstacles.
We find that already  
the presence of few obstacles 
significantly alters 
the collective dynamics 
by creating regions where clogs, intermittent and free flow 
coexist. The system mean speed 
decreases by increasing the density of  
flowing particles or obstacles. 
When increasing the particle speed, 
we find an overall decrease of the mean 
speed, and this 
provides an experimental evidence of the 
"faster is slower" (FIS) effect.
Further, we show how to control and reduce 
the formation of clogs by tuning the 
pair interactions between the moving colloids.
These findings remark the importance of 
particle speed and interactions on the clogging process. 

We transport paramagnetic colloidal particles 
with diameter $d_m=2.8 \rm{\mu m}$ 
by using a magnetic ratchet effect 
generated at the surface of 
a uniaxial ferrite garnet film (FGF).
The FGF is characterized by a parallel stripe pattern
of ferromagnetic domains with alternating up and down magnetization normal to the substrate, and a
spatial periodicity of $\lambda=2.6 \rm{ \mu m}$
at zero applied field, Fig.1(a).
To introduce quenched disorder in the system, 
we use silica particles with a diameter $d_o=5 \rm{\mu m}$, 
thus larger than the magnetic colloids
and distinguishable from them. 
Previous to the experiments, the particles are  
let sediment on top of the FGF,
and irreversibly attached there 
by screening the electrostatic interactions via 
addition of salt. Afterwards, the substrate is dried and refilled with the 
paramagnetic colloids, see~\cite{EPAPS} for more details.
\begin{figure*}[t]
\begin{center}
\includegraphics[width=\textwidth,keepaspectratio]{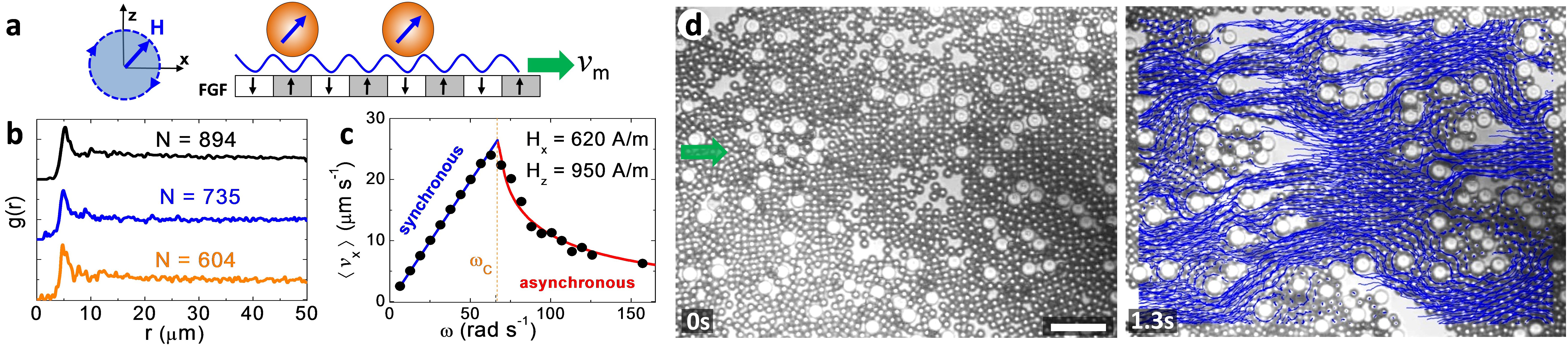}
\caption{
(a) Schematic of the traveling wave potential generated at the surface of the 
FGF when subjected to a rotating magnetic field. 
(b) Pair correlation function $g(r)$ of the silica particles for $3$ samples 
with different number of particles $N$. For clarity the $g(r)$
are rescaled along the $\hat{y}$ axis.
(c) Mean particle speed $\langle v_x \rangle$ 
versus driving frequency $\omega$ in the obstacle-free case. 
Blue [red] fit to the data 
indicates the synchronous, $\langle v_x \rangle = v_m  $ [asynchronous, 
$\langle v_x \rangle= v_m (1-\sqrt{1-(\omega_c/\omega)^2})$] 
regime, being $\omega_c= 66.8\rm{rad s^{-1}}$ the critical frequency. 
(d) Experimental images of a monolayer of  
paramagnetic colloids (black circles) driven through a 
disordered landscape of silica particles (white circles)
towards right ($\hat{x}>0$). 
The particle trajectories are superimposed in blue on the image at the right
($t=1.3\rm{s}$). 
Scale bar is $20\rm{\mu m}$, 
see also MovieS1 in~\cite{EPAPS}.}
\label{figure1}
\end{center}
\end{figure*}
Using this procedure, we generate a random array of obstacles 
evenly distributed across the FGF film. The corresponding
disorder is spatially uncorrelated and has no detectable features 
of ordering, as shown by the absence of secondary peaks in 
the pair correlation functions $g(r)$
of the silica particles, calculated 
for different samples in Fig.1(b). 

The transport mechanism was previously 
introduced for an obstacle-free substrate~\cite{Tie12,Str13},
and here will be briefly described.
Above the FGF, the paramagnetic colloids are 
driven upon application of a rotating magnetic field polarized in the 
$(\hat{x},\hat{z})$ plane,
$\bm{H} = [H_x \cos{(\omega t) \hat{\bm{e}}_x}- H_z \sin{(\omega t) \hat{\bm{e}}_z}]$,
with frequency $\omega$ and amplitudes $(H_x,H_z)$. 
The applied field 
modulates the stray magnetic field at the FGF surface, 
and generates a periodic potential that translates 
at a constant and frequency tunable speed, $v_m=\lambda \omega/(2 \pi)$.
The potential is 
capable of transporting the colloidal particles
that are trapped in its energy minima, Fig.1(a).
As shown in Fig.1(c), 
by varying the driving frequency,
the particles undergo a 
sharp transition from a phase-locked motion with 
$\langle v_x \rangle = v_m$ (synchronous regime),
to a sliding dynamics (asynchronous regime) resulting from the 
loss of synchronization
with the traveling potential.
Throughout this work, we restrict the angular frequency 
to the former regime,
and vary $\omega$ to tune the particle speed.
Given the strong magnetic attraction of the FGF, 
the particle motion is essentially two-dimensional,
with negligible out-of-plane thermal fluctuations. 
Further, the ratio between the field amplitudes $H_x/H_z$
will be used to tune the pair interactions.
However, unless stated otherwise, we initially set this ratio to $H_x/H_z \sim 0.7$
such to minimize these interactions and obtain an 
hard-sphere like behavior. 
  
Fig.1(d) shows two experimental images of a dense collection of 
paramagnetic colloids driven 
against $N=80$ silica particles. From the trajectories, it
emerges that
the magnetic colloids surpass the silica particles  by following a path similar to laminar flow. 
We use an upright microscope equipped with a CCD camera,
to record real-time videos of the system dynamics,
and analyze different subsets of a total field of view of $145 \times 108 \rm{\mu m}^2$. We then 
determine the positions $(x_i(t),y_i(t))$ of each magnetic colloid $i=1...N$ via particle tracking routines~\cite{Cro96}, and measure  
the instantaneous velocity $v_x (t) = \frac{1}{N} \sum_i \frac{dx_i}{dt}$,
and its mean value $\langle v_x (t) \rangle$, 
with the time-average taken in the stationary regime.
Both quantities for Fig.1(d) are shown in~\cite{EPAPS}.

We start by analyzing the particle flow 
at the level of a single  
aperture, 
such as a pair of silica 
particles
separated by a 
surface to surface distance $d \geq 3 \rm{\mu m}$
in a cluster of connected obstacles. 
The complete trapping of the particles 
by the obstacles ($\langle v_x \rangle =0$) is observed for 
lower distances. 
In general, we find  
an intermittent flow 
of the magnetic colloids 
for distance $d\lesssim 4 \rm{\mu m}$, 
when the aperture is normal to the 
driving direction.
Larger aperture between the obstacles,
as the one shown in the inset of Fig.2,
were found to arrest the flow 
only when the vector joining 
the obstacle's centers 
is not perpendicular 
to $\hat{x}$.
The intermittent flow 
arises from the simultaneous arrival
of the paramagnetic colloids 
at the aperture, 
and their accumulation 
in a close packed state, thus locally jammed.
If compared to arches formation in granular systems~\cite{To01,Loz12},
we find that the combined effect of 
hydrodynamic lubrication between the particles
and the oscillations induced by the magnetic ratchet  
make these states rather fragile. 
We analyze different cases, and in all of them 
we measure the distribution $P(t>t_p)$ of time lapse $t>t_p$ 
between the particles passing through the aperture, 
as shown in Fig.2.  
For all driving frequencies explored, 
we find that
$P$ displays a power law tail, with an exponential 
distribution of burst size $S$.
\begin{figure}[t]
\begin{center}
\includegraphics[width=\columnwidth,keepaspectratio]{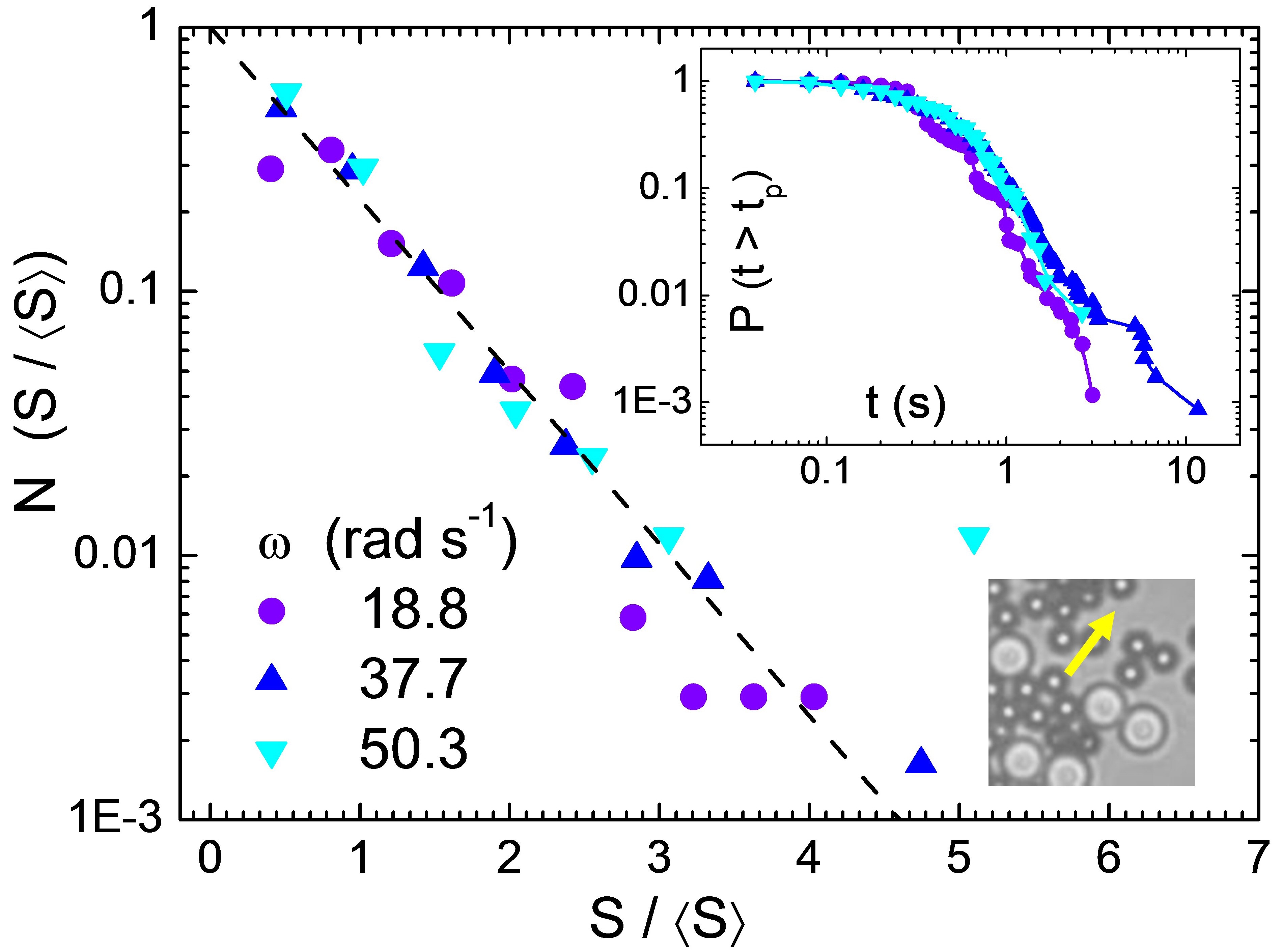}
\caption{Histograms of burst size
$S$ rescaled with respect to $\langle S \rangle$
and measured for different driving frequencies $\omega$.
The dashed line is an exponential fit to the data as guide to the eye. The lower temporal limit used to define the bursts is $t_p= 0.4 \rm{s}$,
the average burst sizes are $\langle S \rangle=2.5$ for $\omega=18.8 \rm{rad s^{-1}}$, $\langle S \rangle=1.9$  for $\omega=37.7 \rm{rad s^{-1}}$
and $\langle S \rangle=1.7$ for $\omega=50.3 \rm{rad s^{-1}}$.
All data are taken for an applied field with $H_x=620 \rm{A m^{-1}}$, 
$H_z=950 \rm{A m^{-1}}$. Top inset: The complementary cumulative distribution function $P$ 
of time lapses $t>t_p$  for magnetic particles passing through two silica obstacles. The small microscope image at the bottom shows 
the considered aperture.}
\label{figure2}
\end{center}
\end{figure}
These observations confirm the universal feature 
of $P$, as similar results were  
demonstrated in the past with sheep, granular particles and 
numerical simulations of sedimenting colloids~\cite{Ike14}. 

Next, we analyze the collective dynamics across the whole landscape 
by varying the area fractions of magnetic colloids, $\Phi_m$ and obstacles, $\Phi_{o}$.
Here $\Phi_j= N_j \pi (d_j/2)^2/A$, being $N_j$ the number of elements $j$
having diameter $d_j$ and located in the area $A$. 
In Fig.3(a) we show the
different dynamic phases in the 
$(\Phi_m,\Phi_o)$ plane 
in terms of the
normalized  
mean speed, 
thus $\langle v_x 	\rangle /v_m \in [0,1]$. 
The particle flow is maximal 
at low obstacle density, $\Phi_o < 0.07$, where few 
silica particles  
are unable to arrest the motion 
of magnetic colloids.
\begin{figure}[b]
\begin{center}
\includegraphics[width=\columnwidth,keepaspectratio]{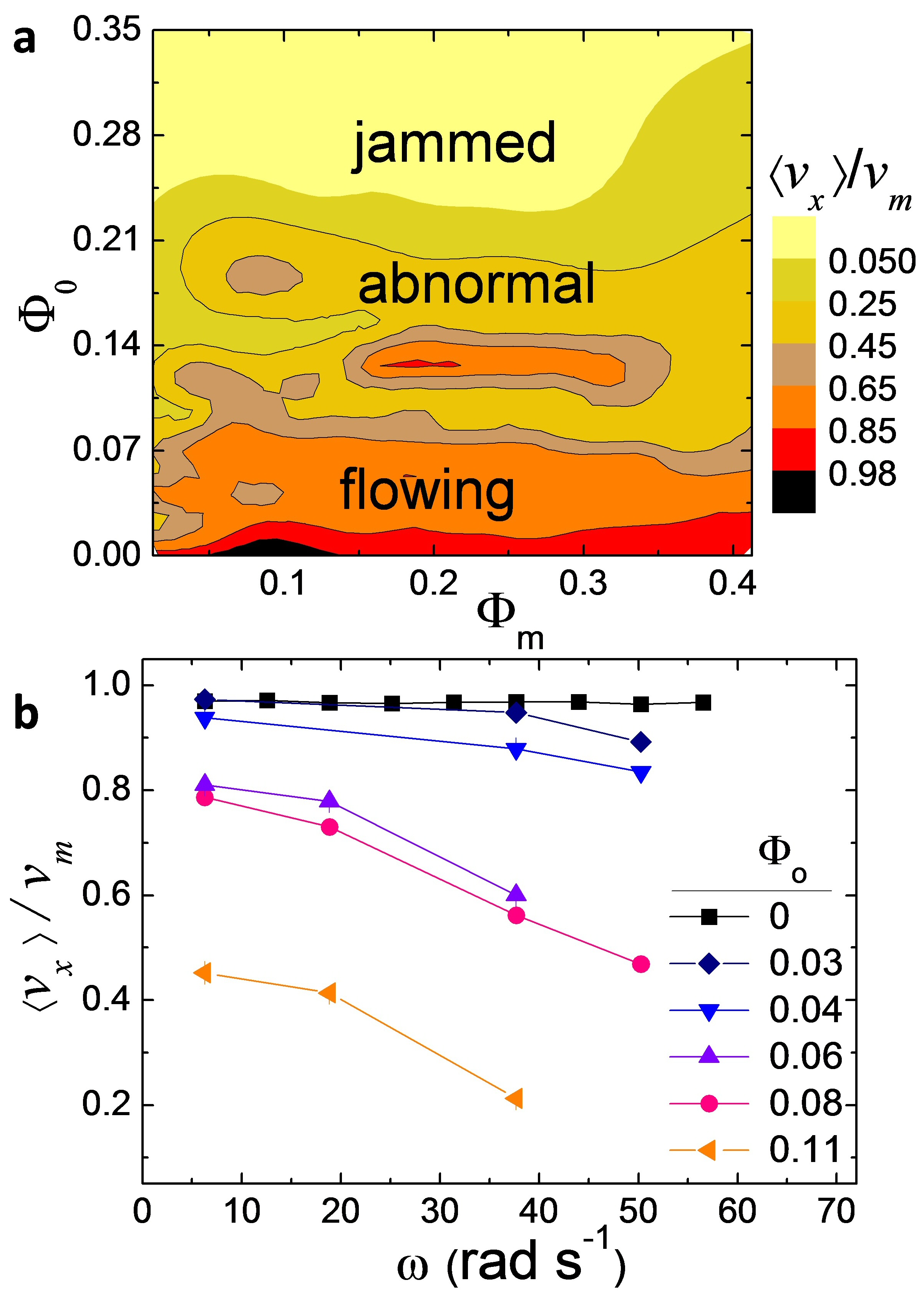}
\caption{(a) Diagram in the $(\Phi_m, \Phi_o)$
plane illustrating regions of normal flow ($\langle v_x \rangle /v_m >0.85$),
abnormal flow characterized by coexistence of clogs and flowing particles  
($\langle v_x \rangle/v_m \lesssim 0.85$), and 
jamming that was found below $\Phi_o\sim 0.25$. 
The magnetic particles are driven in the synchronous regime 
with $\omega=37.7 \rm{rad s^{-1}}$.
(b) "Faster is slower" effect on a disordered landscape: Normalized 
mean velocity $\langle v_x \rangle/v_m$ 
versus  
driving frequency $\omega$ for different disorder densities $\Phi_o$. 
All data in (a) and (b) where 
taken for $H_x= 620\rm{A m^{-1}}$, 
and $H_z= 950\rm{A m^{-1}}$.}
\label{figure3}
\end{center}
\end{figure}
However, we find a decrease 
of the mean speed 
$\langle v_x \rangle/v_m\sim 0.25$ 
on the left, lower corner of the diagram 
($\Phi_m<0.1,\Phi_o<0.07$).
In this situation, the 
magnetic colloids  
that hit the obstacles at the center are not scattered, 
but remain trapped there as long as other particles do not release them via collision.
\begin{figure*}[t]
\begin{center}
\includegraphics[width=\textwidth,keepaspectratio]{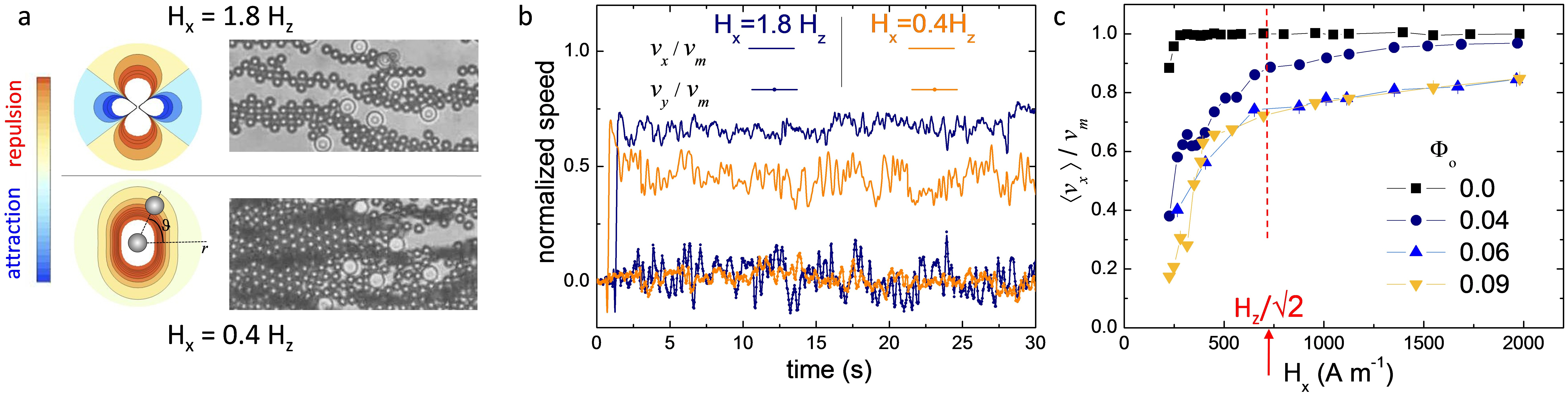}
\caption{(a) Left: Polar plots of the dipolar interactions between two 
paramagnetic colloids calculated 
for $H_x=1.8 H_z$ (top) and $H_x=0.4 H_z$ (bottom), with $H_z=950 \rm{A m^{-1}}$. 
Red (blue) region denotes repulsion (attraction).
Right: Experimental snapshots of the corresponding situations 
realized with $\omega = 37.7 \rm{rad s^{-1}}$,
$\Phi_m=0.3$, $\Phi_o=0.06$, see also MovieS3 in~\cite{EPAPS}. 
(b) Normalized particle velocities ($v_x,v_y$) versus time for the two cases. 
(c) Mean velocity $\langle v_x \rangle/v_m$
versus amplitude of the in-plane component $H_x$,
at different disorder densities $\Phi_o$,
with $\omega = 37.7 \rm{rad s^{-1}}$ and $\Phi_m=0.3$.}
\label{figure4}
\end{center}
\end{figure*}
A similar feature  
has been observed recently  
via numerical simulations of driven disks 
across a disorder landscape~\cite{CReArxiv}.
Increasing $\Phi_o$ reduces the system mean speed, 
as now flowing regions $\langle v_x \rangle \sim v_m$,
coexist with clogged ones $\langle v_x \rangle =0$. The
latter states however, may be easily unclogged by inverting the particle flux, 
that corresponds to reversing the sense of rotation of the applied field, 
$H_x \leftrightarrow -H_x$, see Fig.2 and MovieS2 in~\cite{EPAPS}.
This extended region of "abnormal flow"
includes also cases where the system was almost 
completely arrested $\langle v_x \rangle/v_m \sim 0.3$, 
but it could be again fluidized
by reversing the 
applied field.
At larger obstacle density,  $\Phi_o>0.25$,
the silica particles percolate through the system and surround 
the magnetic colloids, thus impeding any net movement.  
Here the system forms a 
solid-like or "jammed" phase from the very beginning, 
and cannot be refluidized 
by the applied drive, see  
Fig.3 and MovieS4 in~\cite{EPAPS}. 
Thus, in our system we never observe fully clogged states, 
where  the mean speed vanishes after a transitory period,
and that can be refluidized by inverting the particle current
\footnote{We did not explore
the high dense case of magnetic colloids, $\Phi_m>0.4$,
as there we found that strong dipolar interactions 
favor particle jumping across the obstacles 
leaving the two dimensional confinement.}.
In contrast, we observe 
that  the presence of few obstacles 
strongly reduces the jamming threshold
where $\langle v_x \rangle=0$ from the very beginning. 
In absence of silica particles, 
the driven colloids will jam near the close packing density, 
$\Phi_m=\pi/2 \sqrt{3}\sim 0.9$.
\footnote{Experimental limitation impeded us 
to reach this value, we find that the system 
was observed to flow as a solid like material
up to $\Phi_m\sim 0.8$.} 
A small concentration of obstacles $\Phi_o = 0.25$ reduces the jamming 
threshold to $\Phi_m= 0.3$ or below. 

In Fig.3(b) we investigate the effect of varying 
the driving frequency $\omega$, and thus the particle speed
on the system dynamics, by keeping constant $\Phi_m=0.25$. 
We experimentally observe  
the "faster is slower" (FIS) effect, 
where the mean speed of the system decreases as  
$\omega$ increases, i.e. the single particle velocity in the obstacle free case. 
The FIS effect was originally observed when simulating 
the dynamics of pedestrians trying to escape 
through a narrow exit~\cite{Hel00}. A recent surge of interest 
on FIS~\cite{Sti17} resulted from the possibility to observe 
a similar effect in other systems~\cite{Pas15},
although its investigation has been limited to a single aperture.  
On a collective level, the effect of the 
velocity drop by an increase 
in the applied driving force 
has been reported by different
theoretical works
in condensed matter systems~\cite{Dha84,Zia02,Jac08,Bai15,Ben16,Rei18},
although with no experimental evidence. 
Here, when considering an heterogeneous landscape, 
we find that FIS is greatly enhanced
by disorder. For example a small area fraction of obstacles $\Phi_o = 0.08$,
is able to reduce the speed of a $\sim 45 \%$ with respect to the free case when increasing the driving frequency 
from $6.3\rm{rad s^{-1}}$ to 
$50.3\rm{rad s^{-1}}$, as shown in 
the magenta curve in Fig.3(b). 

Our system 
further allows to tune the pair interactions between the moving 
colloids by varying the 
amplitude $H_x$ of the in-plane field. 
Above the 
magnetically modulated FGF,
the effective interaction potential between two paramagnetic colloids 
can be calculated
via a time average~\cite{Str14}.
In polar coordinates $(r,\vartheta)$ 
this potential can be written as 
$U_{d}=\alpha[H_x^2(1+3\cos{2\vartheta})-2H_z^2]/r^3$,
where $\alpha=\chi d_m^3 /(96 \lambda^3 M_s^2)$, 
$\chi=0.4$ is the effective volume susceptibility of the 
particles and $M_s=1.3 \cdot 10^4 \rm{A m^{-1}}$ is the saturation magnetization of the FGF. 
The previous equation shows that, at a fixed value of $H_z$,
the sign of the pair interaction is repulsive 
when $H_x<H_z/\sqrt{3\cos{\vartheta}^2-1}$, while being attractive in the other case.
As shown in Fig.4(a), the dipolar interactions between two colloids 
are strongly anisotropic. We consider two representative 
cases at fixed frequency, characterized by    
a complete repulsion for $H_x=0.4 H_z$, and attraction (repulsion),
for $H_x=1.8 H_z$ when the relative angle $\vartheta$ between the particles 
lies in the region 
$\vartheta\in [n\pi-\vartheta_m,n\pi+\vartheta_m]$ (otherwise).
Here $n=0,1...$ and $\vartheta_m=48.7^{o}$
gives the condition for vanishing magnetic interactions.
The corresponding dynamics for these two cases are 
analyzed in Fig.4(b) in terms of the particle velocities 
parallel and normal to the direction of motion, $v_x$ and $v_y$ 
respectively. 
We find that the attractive interactions,
combined with repulsion at large $\vartheta$, facilitate 
the formation of
parallel trains of magnetic colloids that 
are able to easily move across the disorder landscape.
These trains slide plastically through
the obstacles and, as a result,
the amplitude of fluctuations along the 
normal direction 
increases, being the variance 
$\delta v_y = 1.15 \rm{\mu m s^{-1}}$ 
for the attractive case, the double than in the repulsive case ($\delta v_y = 0.54 \rm{\mu m s^{-1}}$). 
This situation may appear, at a first glance, counter intuitive 
since one mechanism of clogging in microchannels 
is the aggregation of attractive particles at the entrance~\cite{Dre17}.
However, here the anisotropy of the 
pair interactions facilitate the system fluidization, as it
induces the formation of elongated structures
with a certain flexibility,
rather than isotropic compact clusters.
When $H_x=0.4H_z$, the driven monolayer 
has a larger inter-particle distance that 
increases $\Phi_m$, reducing the colloidal flow
as shown in Fig.3(a).
We confirm this general trend by reporting 
in Fig.4(c) the mean speed  
for different disorder densities. 
We finally note that for the field strengths employed here, $H_{x,z}\in[0.3,2] \rm{kA m^{-1}}$, the magnetic 
dipolar interactions between two particles at
a distance $d_p$ are of the order  
$U_{d}\in[2,95] k_B T$, thus higher than other 
interactions as electrostatic or thermal ones.

To conclude, we studied the clogging 
process in a system composed of 
an ensemble of interacting paramagnetic colloids
driven across a quenched disorder landscape. 
We have reported a rich phenomenology, including 
the enhancement of the "faster is slower" effect 
by quench disorder and the role 
of the pair interactions on system 
collective transport. 
The understanding and control of 
clogging process along the lines of
this work may be important 
in different technological contexts
and on different length scales. 
A further potential avenue of this work 
may be related to investigating the clogging transition and 
its connection to jamming in pinned systems, 
a recent theoretical hot spot~\cite{Bri13,Gra16,Rei17}.

\begin{acknowledgments}
We thank Tom H. Johansen for the FGF film, H. Massana-Cid and 
F. Martinez-Pedrero for initial experiments,
I. Zuriguel, H. L\"owen and C. Reichhardt
for stimulating/inspiring discussions. 
R. L. S. acknowledges support from the Swiss National Science Foundation grant 
No. 172065. P.T. acknowledges support from the ERC starting grant "DynaMO" (335040)
and from MINECO (FIS2016-78507-C2) and DURSI (2014SGR878).
\end{acknowledgments}
\bibliography{biblio}
\end{document}